\documentstyle[aps,12pt,manuscript]{revtex}
\begin{document}
\preprint{INJE--TP--95--3}
\def\overlay#1#2{\setbox0=\hbox{#1}\setbox1=\hbox to \wd0{\hss #2\hss}#1%
\hskip -2\wd0\copy1}

\title{An infinite number of potentials surrounding 2d black hole}

\author{ Y. S. Myung }
\address{Department of Physics, Inje University, Kimhae 621-749, Korea}

\maketitle

\vskip 2in

\begin{abstract}
We found an infinite number of potentials surrounding 2d black hole.
 According to the transmission ${\cal T}$ and  reflection ${\cal R}$
coefficients for scattering of string fields off 2d black hole,
we can classify an infinite number of potentials
into three : graviton-dilaton, tachyon and the other types.
We suggest that the discrete states from all the Virasoro levels   be
candidates for new
potentials (modes).
\end{abstract}

\newpage

It is always possible to visualize the
black hole as presenting an effective potential barrier (or well) to the
on-coming waves [1].
In the case of 4d Schwarzschild black hole, two kinds
of Schr\"odinger type equation arose from the metric perturbations.
One is the Regge-Wheeler (RW) equation in the axial (odd-parity)
perturbation [2],

\begin{equation}
{d^2 \Psi_o \over d r^{*2} } + (k^2 - V_{RW}) \Psi_o = 0,
\end{equation}
Here units are used in which $G = c  = 2 M_4 = 1, r^* = r + \ln(r-1) $, so
that the horizon is at
$r^* = - \infty$ ($r=1$).
The RW potential $V_{RW}$ is given by
$$ V_{RW} = { 2 (n+1) r - 3 \over r^4 } (r -1)   $$
with  $n = (l -1)(l+2)/2, l\geq 2$.
The other is  the Zerilli equation
\begin{equation}
{d^2 \Psi_e \over d r^{*2} } + (k^2 - V_Z) \Psi_e = 0.
\end{equation}
which differs only in the details of the potential
$$ V_Z = { 2 (n+1) r^3 + 3 r^2 + 9r/2n + 9/4n^2  \over
           r^4 ( r+ 3/2n)^2  } (r -1).   $$
The Zerilli equation arose in the study of polar (even-parity)
perturbations in the same
formalism.
Although these potentials have different forms, they  are equivalent.
Chandrasekhar have showed that $V_{RW}$ and $V_Z$ are equivalent in the
sense of producing
the same reflection $({\cal R})$ and
transmission $({\cal T})$ coefficients [1,3].
Further, Anderson and Price showed that there may exist an infinite number
of equivalent potentials surrounding the 4d real black holes [4].
In the Schwarzschild black holes $V_{RW}$ and $V_Z$
are considered as only two realizations of an infinite number of possible
potentials.
Four potentials are realized for four graviton-Maxwell modes in the
Reissner-Nordstr\"om black hole.
However, the finding of an infinite number of potentials is a difficult
problem in the real black
holes. Moreover it is not clear what  the candidates for new potentials
(modes) are.
 We  here consider a simpler toy model, 2d stringy black hole in which an
analogous problem
can be imposed and exactly solved [5,6].
The 2d dilaton gravity is far from being 4d realistic models in the
sense that two propagating gravitons are missing.

In this paper we will find  an infinite number of potentials in the 2d
black hole.
In analyzing  two-dimensional stringy black hole,
we begin with one graviton $(h)$, one dilaton $(\phi)$ and one tachyon $(t)$.
These are, in turn, combined into one graviton-dilaton $(h - \varphi)$, the
other
$(h + \varphi)$ and tachyonic ($t$) modes [7]. All of these modes satisfy
the Schr\"odinger
type equation (14). According to the transmission and reflection coefficients,
we can classify an infinite number of potentials
into three : graviton-dilaton, tachyon and the other types.
Finally, we discuss the candidates for  the new potentials (modes)
except graviton, dilaton and tachyon.

The $\sigma$-model action of 2d critical string theory for graviton
($g_{\mu \nu}$), dilaton ($\Phi$), and tachyon ($T$) ($\mu, \nu = 0, 1$)
is given by [8]

\begin{equation}
S_\sigma = {1 \over 8 \pi \alpha^\prime} \int d^2 z \sqrt{G}
[ g_{\mu \nu}(x) \nabla x^\mu \nabla x^\nu + \alpha^\prime R \Phi(x) + 2T(x) ]
\end{equation}
with  $x^0 = \theta, x^1 = \phi $. The Minkowiski signature is recovered by the
analytic continuation of $\theta = i \tau.$
 The conformal invariance of $S_{\sigma}$ requires the following
$\beta$-function equations
to be satisfied :

\begin{eqnarray}
&&R_{\mu\nu} + \nabla_\mu \nabla_\nu \Phi + \nabla_\mu T \nabla_\nu T = 0,  \\
&&R + (\nabla \Phi)^2 + 2 \nabla^2 \Phi + (\nabla T)^2 - 2 T^2 - 8 = 0,  \\
&&\nabla^2 T + \nabla \Phi \nabla T + 2 T = 0.
\end{eqnarray}
These equations can also be derived from the 2d target space effective
action in [9] with the substitutions

\begin{equation}
-2\Phi_{DL} \rightarrow \Phi, T_{DL} \rightarrow T, -R_{DL} \rightarrow R.
\end{equation}

Let us begin with the static background solutions of
the graviton-dilaton sector without the tachyonic condensation,

\begin{equation}
\bar \Phi = 2 Q \phi,~~~ \bar T = 0,
{}~~~ \bar g_{\mu\nu} =
 \left(  \begin{array}{cc} - f & 0  \\
                             0 & f^{-1}   \end{array}   \right),
\end{equation}
where

\begin{equation}
f = 1 - M e^{- 2 Q \phi}, ~~~~~~ Q = \sqrt 2 .
\end{equation}
Here the parameter $M$ ($>0$) is proportional to the mass of the black hole.
All these solutions approach the linear dilaton vacuum in the asymptotically
flat region ($\phi \to + \infty$).  The event horizon of the black hole
occurs at
$\phi_{EH} = {1 \over 2 \sqrt 2} \ln M$. We choose $M=1$ and
$\phi_{EH} = 0$ for simplicity.  As a consequence,  $\phi_{EH} = 0$ defines
a null surface. Since the region interior to the horizon ($\phi < 0$) is of
no relevance to our consideration,  the new coordinate
($\phi^*$) is introduced as

\begin{equation}
\phi^* \equiv \phi + {1 \over 2 \sqrt 2} \ln (1 - e^{- 2 \sqrt 2 \phi}).
\end{equation}
Note that $\phi^*$ ranges from $- \infty$ to $+ \infty$, while $\phi$ ranges
from the event horizon of the black hole ($\phi_{EH} = 0$) to $+ \infty$.

To study the scatterings off black hole, let us introduce the small
perturbed fields  $h_{\mu\nu} (\phi,\tau)$, $\varphi
(\phi,\tau)$ and $t(\phi,\tau)$ as [7]

\begin{eqnarray}
g_{\mu\nu} &=& \bar g_{\mu\nu} + h_{\mu\nu}
            = \bar g_{\mu\nu} [1 - h (\phi,\tau)],   \\
\Phi &=& \bar \Phi + \varphi (\phi,\tau),   \\
T &=&\bar T + \tilde t \equiv \exp (-{\bar \Phi \over 2}) [ 0 + t
(\phi,\tau) ].
\end{eqnarray}
Substituting (11)-(13) into  (4)-(6) and then keeping up to linear order of
perturbed
fields, one gets the linearized equations for  $\Psi = h - \varphi, h +
\varphi, t$ modes.

Considering   $\Psi( \phi^*, \tau ) = \psi ( \phi^* ) e^{-i \omega \tau}$,
we find the Schr\"odinger type equation for 2d black hole
\begin{equation}
[{d^2 \over d \phi^{*2}}+\omega^2 -{V_0 \over (\cosh \sqrt{2} \phi^*)^2 }]
\psi = 0.
\end{equation}
Here $V_0 =0$  for ($h - \varphi$) , $V_0 = -4$ for ($h + \varphi$), and
$V_0 = 1/2$ for ($t$).
The above equation governs the propagation
of  all perturbing string fields in the black hole background.
In order to solve the equation (14), we
make the  substitution
\begin{equation}
\psi = (\cosh \sqrt{2} \phi^*)^{-2 \lambda} u
\end{equation}
with
\begin{equation}
\lambda = {1 \over 4} ( \sqrt{1 - 2 V_0} - 1 ).
\end{equation}
Then the equation for $u$ takes the form
\begin{equation}
{d^2 u \over d \phi^{*2}}
- 4 \sqrt{2} \lambda \tanh \sqrt{2} \phi^* {d u \over d \phi^*}
+ 8 ( \lambda^2 + {1 \over 8} \omega^2 ) u = 0.
\end{equation}
Two kinds of solutions to the Schr\"odinger equation (14) correspond to
scattering and bound states.
Here we consider only the case of $\omega^2 > 0$ for scattering state.
 The other case ($\omega^2 < 0$)
 is used to study the bound state problem.
If we introduce a new independent variable

$$z = - ( \sinh \sqrt{2} \phi^*)^2,$$
then the equation for $u$ reduces to the hypergeometric equation

\begin{equation}
z (1 -z) {d^2 u \over d z^2} + [{1 \over 2} - (1 - 2 \lambda) z ]{d u \over d
z}
- ( \lambda^2 + {1 \over 8} \omega^2 ) u = 0.
\end{equation}
The parameters $\alpha$, $\beta$, $\gamma$ which occur in the general form
of the hypergeometric equation,

\begin{equation}
z (1 -z) {d^2 u \over d z^2} + [\gamma - (\alpha + \beta + 1) z ]{d u \over d
z}
- \alpha \beta u = 0,
\end{equation}
take in our case the following values :

$$\gamma = {1 \over 2}, ~~~~ \alpha = - \lambda + { i \omega \over 2
\sqrt{2} },
{}~~~~ \beta  = - \lambda - { i \omega \over 2 \sqrt{2} }.  $$
 Two exact solutions of equation (18) are of the forms

\begin{eqnarray}
u_1 &=& F( - \lambda + { i \omega \over 2 \sqrt{2} },
 - \lambda - { i \omega \over 2 \sqrt{2} }, {1 \over 2} ; z),  \\
u_2 &=& \sqrt{z} F( - \lambda + { i \omega \over 2 \sqrt{2} } + {1 \over 2},
 - \lambda - { i \omega \over 2 \sqrt{2} } + {1 \over 2} , {3 \over 2} ; z).
\end{eqnarray}
The general form of the wavefunction is

\begin{eqnarray}
\psi &=& C_1 (\cosh\sqrt{2} \phi^*)^{- 2 \lambda}
 F( - \lambda + { i \omega \over 2 \sqrt{2} },
 - \lambda - { i \omega \over 2 \sqrt{2} }, {1 \over 2} ; z)  \nonumber \\
  &+& C_2 (\cosh\sqrt{2} \phi^*)^{- 2 \lambda} \sqrt{z}
F( - \lambda + { i \omega \over 2 \sqrt{2} } + {1 \over 2},
 - \lambda - { i \omega \over 2 \sqrt{2} } + {1 \over 2} , {3 \over 2} ; z).
\end{eqnarray}
Here the coefficients $C_1$ and $C_2$ will be determined by requiring the
boundary conditions
at $\phi^* \to \pm \infty$ [10].

For $V_0>{1 \over 2}$, we have
\begin{eqnarray}
{\cal T} &=&
{ ( \sinh (\pi \omega / \sqrt{2}) )^2
\over ( \sinh (\pi \omega / \sqrt{2}) )^2 + ( \cosh ({\pi \over 2} \sqrt
{2V_0 -1}))^2},\\
{\cal R} &=&
{ ( \cosh({\pi \over 2} \sqrt {2V_0 -1}) )^2
\over ( \sinh (\pi \omega / \sqrt{2}) )^2  +  ( \cosh({\pi\over2} \sqrt
{2V_0 -1}))^2}.
\end{eqnarray}
In this case, one cannot obtain an infinite number of potentials because of
the non-periodic nature
of the hyperbolic functions.

For $V_0 \leq {1 \over 2}$, we obtain the transmission coefficient
\begin{equation}
{\cal T} =
{ ( \sinh (\pi \omega / \sqrt{2}) )^2
\over ( \sinh (\pi \omega / \sqrt{2}) )^2 + ( \cos \pi( 2 \lambda +{1 \over
2}))^2}.
\end{equation}
Similarly one can derive the reflection coefficient
\begin{equation}
{\cal R}=
{ ( \cos \pi( 2 \lambda +{1 \over 2}))^2
\over ( \sinh (\pi \omega / \sqrt{2}) )^2  +  ( \cos \pi( 2 \lambda +{1
\over 2}))^2}.
\end{equation}
We classify  these forms   into  the following types.

\subsection{graviton-dilaton type : $2 \lambda = n$}

This comes from the condition : $\cos \pi( 2 \lambda +{1 \over 2})=0$.
Using the relation (16), we obtain an infinite series
\begin{equation}
V_0=0 (n=0),~~V_0=-4 (n=1),~~V_0=-12 (n=2),~~V_0=-24 (n=3),\cdots,
\end{equation}
which give us the same transmission ( ${\cal T}^{g-d}$) and reflection
(${\cal R}^{g-d}$) coefficients.
Note that $ h + \varphi$ ($ h - \varphi$) recover from this series  when  $
n=0 (n=1)$ respectively.
The transmission  coefficient is given by
\begin{equation}
{\cal T}^{g-d} = |T^{g-d}|^2 = 1
\end{equation}
This means that there is no reflection, i.e. ${\cal R}^{g-d}=|R^{g-d}|^2  = 0$.
Even though the different potential wells have
arisen from the black hole,  all modes which belong to this series
propagate freely from  $+ \infty$ to $- \infty$.
This corresponds to the  total transmission [11].
For example, there is anomalously large transmission in the low energy
electrons (0.1 eV) scatterings
off noble gases such as neon or argon. For the electron scatterings
 the prototype potential is the square well,
instead of $-{1 \over \cosh^2x}$.  This type of scattering can thus  be
understood by noting
the analogies : graviton-dilaton modes $\iff$ electrons, 2d black hole
$\iff$ neon or argon.

\subsection{tachyon type : $2\lambda= n-1/2$}
Requiring that $\cos \pi( 2 \lambda +{1 \over 2})= \pm 1$,
we obtain the following infinite series
\begin{equation}
V_0={1 \over  2} (n=0),~~V_0=-{3 \over  2}(n=1),~~V_0=-{15 \over  2}
(n=2),~~V_0=-{35 \over  2} (n=3)
,\cdots.
\end{equation}
The first one corresponds to the tachyon mode.
All of these lead to the same transmission ( ${\cal T}^t$) and reflection
(${\cal R}^t$) coefficients

\begin{eqnarray}
{\cal T}^t &=&
{ ( \sinh (\pi \omega / \sqrt{2}) )^2  \over  1 + ( \sinh (\pi \omega /
\sqrt{2}) )^2 },\\
{\cal R}^t &=& {1 \over  1 + ( \sinh (\pi \omega / \sqrt{2}) )^2 }.
\end{eqnarray}
As might be expected, one finds that ${\cal T}^t + {\cal R}^t = 1$.
Given the energy of mode $(E=\omega^2)$, this case has the maximum
reflection and minimum
transmission.

\subsection{the other types}

For $0<|\cos \pi( 2 \lambda +{1 \over 2})|<1$, we have the transmission and
reflection coefficients
 which belong to
\begin{equation}
{\cal T}^t <{\cal T} <1,~~~~ 0<{\cal R} <{\cal R}^t.
\end{equation}
Here we can find  an infinite number of potentials which give us the same
physical
consequences.

Now let us discuss our results.
The number of degrees of freedom for the gravitational field ($h_{\mu\nu}$) in
$d$-dimensions is [12]
$$ {1 \over 2} d (d +1) - 1 - d - (d-1) = {1 \over 2} d (d -3).  $$
For 4d Schwarzschild case, we obtain two propagating physical gravitons
(RW in (1) and Zerilli in (2) cases).
There is no two dimensional analog of the Schwarzschild solution since the
Einstein equation
is trivial  for d=2 ($R_{\mu\nu} - {1 \over 2}g_{\mu\nu}R=0$).
Also this counting is $-1$ for $d = 2$.
This means that in two dimensions, the contribution of graviton is equal and
opposite to that of a spinless particle (dilaton).
We recognize from (11) that the gauge-fixing for graviton is lack  on the
basis of 4d gravity theory.
The combined modes of graviton and dilaton ($h - \varphi, h + \varphi$)
should have zero physical degree of freedom [13].
Even though the potential well for $h-\varphi$ has
arisen from the black hole,  this mode can be removed by the coordinate
transformation (translation) [14].
In view of the above, it is obvious that two graviton-dilaton modes
cannot be realized into the physically propagating modes.
Thus the net physical degrees of freedom should be given by
$$-1 (graviton) + 1 (dilaton) + 1 (tachyon) = 1 (tachyon).$$
The above implies that the tachyon  is only a physical degree of
freedom. If additional modes with  $2\lambda= n-1/2$ except tachyon may
exist, they
are all the physical degrees of freedom.
It is  clear that  the  unexplored modes with  $2\lambda=n$ except
two graviton-dilaton modes are the physical degrees of freedom.
For example, we consider the CGHS model with $N$ conformal matters ($f_i$)
[15] instead of tachyon.
Although  all of these belong to the $2\lambda=n$ case ($V_0=0, n=0$), they
correspond to the
physical fields. This is because $f_i$ are external matter fields.
By the similar way it is also suggested that all external modes including
the other
types correspond to the physical degrees of freedom.
In the 2d dilaton black hole, the modes ($h - \varphi, h + \varphi, t$) are
only three
realizations of an infinite number of modes. Among these the tachyon
(external matter) turns out
to be a physically propagating one.
By the similar way, the modes ($h - \varphi, h + \varphi, f_i$) in CGHS
model are the $(2 + N)$
realizations of an infinite number of modes. Here $f_i$ are physical modes.

What is the origin of the infinite number of new  potentials (equations)?
In order to answer to this question, one has to recognize that the key
equations in (14)
are derived from the string perturbations propagating under the 2d black hole.
 The new equations may come from another
string fields except graviton, dilaton and tachyon.  There are an infinite
number
of physical states
in the string theories, the discrete states arising from all the Virasoro
levels including
graviton-dilaton sector [16]. However, having application of string
theories to 2d black hole, two
graviton-dilatons turn out to be nonpropagating modes. Thus  the remaining
discrete states from all
the Virasoro levels  may be  candidates for new potentials (modes)
surrounding 2d black hole.

\acknowledgments
This work was supported in part by NONDIRECTED RESEARCH FUND, Korea
Research Foundation, 1994.


\newpage

\begin{references}
\bibitem{Chandra1} S. Chandrasekhar,  {\it The Mathematical Theory of Black
Hole}

(Oxford
Univ. Press, New York, 1983).
\bibitem{Regge} T. Regge and J. A. Wheeler, Phys. Rev. {\bf 108} (1957) 1403 ;
           C. V. Vishveshwara, Phys. Rev. {\bf D1} (1970) 2870;
           F. J. Zerilli,  Phys. Rev. Lett. {\bf 24} (1970) 737.
\bibitem{Chandra2} S. Chandrasekhar, in {\it Space and Geometry}, eds.
        R. A. Matzner and L. C. Shepley (Texas Univ. Press, Austin, 1982)
p.120.
\bibitem{And1} A. Anderson and R. H. Price, Phys. Rev. {\bf D43} (1991) 3147.
\bibitem{Witten} E. Witten, Phys. Rev. {\bf D44} (1991) 314 ;
                G. Mandal, A. Sengupta and S. R. Wadia, Mod. Phys. Lett.
{\bf A6} (1991) 1685.
\bibitem{Harv} J. A. Harvey and A. Strominger, in {\it String theory and
quantum gravity '92} :
               Proc. the Trieste Spring School {\&} Workshop (ICTP, March 1992)
               eds. J. Harvey, {\it et al} (World Scientific, Singapore,
1993) p.122.
\bibitem{Kim1}  J. Y. Kim, H. W. Lee and Y. S. Myung, Phys. Lett. {\bf
B328} (1994) 291;
               ; Y. S. Myung, Phys. Lett. {\bf B334} (1994) 29;
                Y. S. Myung, J.Y. Kim, C. Jue, Phys. Lett. {\bf B341}
(1995) 273.
\bibitem{Rama} S. K. Rama, Phys. Rev. Lett. {\bf 70} (1993) 3186;
               K. Ghoroku, Phys. Lett. {\bf B347}  (1995) 21.
\bibitem{deAlwis} S. P. deAlwis and J. Lykken, Phys. Lett. {\bf B269}
(1991) 464.
\bibitem{Kim2} J. Y. Kim, H. W. Lee and Y. S. Myung, Phys. Rev. {\bf D50}
(1994) 3942.
\bibitem{Gasi} S. Gasiorowicz, {\it Quantum Physics} (Wiley, New York,
1874) p.80.
\bibitem{Weinberg} S. Weinberg, in {\it General relativity}, eds.
        S. W. Hawking and W. Israel (Cambridge Univ. Press, London, 1979)
p.790.
\bibitem{Horo} G. T. Horowitz, p. 80 in Ref[3].
\bibitem{Kim}  W. T. Kim, J. Lee and Y. J. Park, Phys. Lett. {\bf B347}
(1995) 217.
\bibitem{Callan} C. G. Callan, S. B. Giddings, J. A. Harvey and A. Strominger,
                                                                Phys. Rev.
{\bf D45} (1992) R1005.
\bibitem{Witten} M. B. Green, J. H. Schwarz and E. Witten,
                 {\it Superstring Theory, Vol.1}

(Cambrid
ge  Univ. Press, New York, 1987)




\end{references}
\end{document}